\documentclass[doublecolumn]{aa}
\usepackage[utf8]{inputenc}
\usepackage{graphicx}
\usepackage[english]{babel}
\usepackage{natbib,twoopt}
\usepackage[breaklinks=true]{hyperref} 
\bibpunct{(}{)}{;}{a}{}{,}             
\usepackage{amsmath}
\usepackage{mathtools}
\usepackage{amssymb}
\usepackage{txfonts}
\usepackage{color}
\usepackage{rotating}
\usepackage{enumitem}
\usepackage[labelfont=bf]{caption}
\usepackage{multirow}
\usepackage{booktabs}
\usepackage{subcaption}
\usepackage{ragged2e}
\usepackage{threeparttable}
\usepackage{placeins}
\usepackage{url}
\hypersetup{colorlinks=true,citecolor=blue}
\usepackage{footnotebackref}

\newcommand{\eq}[1]{Eq.\,(\ref{#1})}

\newcommand{\fig}[1]{Fig.\,\ref{#1}}

\newcommand{\sect}[1]{Sect.\,\ref{#1}}

\newcommand{\app}[1]{Appendix\,\ref{#1}}
\newcommand{\tab}[1]{Table\,\ref{#1}}
\newcommand{\tabs}[2]{Tables\,\ref{#1} and \ref{#2}}

\begin{document}

\title{SpeCT: A state-of-the-art tool to calculate correlated-$k$ tables and continua of CO$_2$-H$_2$O-N$_2$ gas mixtures}
\author{Guillaume Chaverot\inst{1,2}, Martin Turbet\inst{3,4}, Ha Tran\inst{3}, Jean-Michel Hartmann\inst{3}, Alain Campargue\inst{5}, Didier Mondelain\inst{5} Emeline Bolmont\inst{2,6}}

\offprints{G. Chaverot,\\ email: guillaume.chaverot@univ-grenoble-alpes.fr}

\institute{
$^1$ Univ. Grenoble Alpes, CNRS, IPAG, 38000 Grenoble, France\\
$^2$ Life in the Universe Center, Geneva, Switzerland\\
$^3$ Laboratoire de M\'et\'eorologie Dynamique/IPSL, CNRS, Sorbonne Universit\'e, \'Ecole Normale Sup\'erieure, PSL Research University, \'Ecole Polytechnique, Institut Poytechnique de Paris, 75005 Paris, France\\
$^4$ Laboratoire d'astrophysique de Bordeaux, Univ. Bordeaux, CNRS, B18N, allée Geoffroy Saint-Hilaire, 33615 Pessac, France\\
$^5$ Univ. Grenoble Alpes, CNRS, LIPhy, 38000 Grenoble, France\\
$^6$ Observatoire astronomique de l'Universit\'e de Gen\`eve, Chemin Pegasi 51, CH-1290 Versoix, Switzerland\\}

  \date{Accepted 18 August 2025}
  \abstract{
   A key challenge in modeling (exo)planetary atmospheres lies in generating extensive opacity datasets that cover the wide variety of possible atmospheric composition, pressure, and temperature conditions. This critical step requires specific knowledge and can be considerably time-consuming. To circumvent this issue, most available codes approximate the total opacity by summing the contributions of individual molecular species during the radiative transfer calculation. This approach neglects inter-species interactions, which can be an issue for precisely estimating the climate of planets.  

   To produce accurate opacity data, such as correlated-$k$ tables, $\chi$ factor corrections of the far-wings of the line profile are required. We propose an update of the $\chi$ factors of CO$_2$ absorption lines that are relevant for terrestrial planets (pure CO$_2$, CO$_2$-N$_2$ and CO$_2$-H$_2$O). These new factors are already implemented in an original user-friendly open-source tool designed to calculate high resolution spectra, named \texttt{SpeCT}. The latter enables to produce correlated-$k$ tables for mixtures made of H$_2$O, CO$_2$ and N$_2$, accounting for inter-species broadening. 
   In order to facilitate future updates of these $\chi$ factors, we also provide a review of all the relevant laboratory measurements available in the literature for the considered mixtures.    
   Finally, we provide in this work 8 different correlated-$k$ tables and continua for pure CO$_2$, CO$_2$-N$_2$, CO$_2$-H$_2$O and CO$_2$-H$_2$O-N$_2$ mixtures based on the MT\_CKD formalism (for H$_2$O), and calculated using \texttt{SpeCT}. These opacity data can be used to study various planets and atmospheric conditions, such as Earth's paleo-climates, Mars, Venus, Magma ocean exoplanets, telluric exoplanets.
   }
  

\titlerunning{SpeCT: Spectra for correlated-$k$ Tables}
\authorrunning{Chaverot et al.} 

   \maketitle

\section{Introduction}

The continuous improvement of remote sensing instruments and methods leads to the study of smaller and smaller planets, toward Earth-sized rocky exoplanets. Along this path, several major space missions aim to detect and characterize small rocky targets, potentially hosting liquid water. 
The James Webb Space Telescope (JWST) is already observing planetary atmospheres, of the TRAPPIST-1 system for instance, giving information on their composition with an unprecedented precision \citep{greene_thermal_2023, zieba_no_2023}.
This effort will be reinforced by the VLT, and especially by the ANDES instrument \citep{marconi_andes_2022}, planned for a first light in 2030, which will combine transmitted and reflected lights to characterize rocky planets around M-dwarfs. 
In a near future, PLATO \citep{rauer_plato_2025}, planned for a launch in 2026, will be dedicated to the detection of exoplanets orbiting FGK type stars in their Habitable Zone (HZ, \citealt{kasting_habitable_1993, kopparapu_habitable_2013}). On a longer timescale, the Habitable Worlds Observatory (HWO) and the Large Interferometer for Exoplanets (LIFE, \citealt{quanz_large_2022}) aim to detect biosignatures in the atmosphere of a very few planetary candidates orbiting solar-type stars. 

As the information provided by these instruments is coming from the atmosphere of the planet, climate modeling is - and will be - a key scientific step to understand other worlds. For every type of climate model (from 1-D to 3-D), one of the major processes is the radiative transfer, which requires opacity data. For the most complex models, such as 3-D Global Climate Models (GCMs), the computation of the radiative transfer is often the most time-consuming step. The calculation of the exact absorption spectra, at each time-step and in every cell of the model grid is thus done exceptionally, for specific cases \citep[e.g.][]{ding_new_2019}.

A compromise is thus required between accuracy and efficiency when creating opacity data dedicated to these models. Line-by-line calculations are time-consuming while gray opacities (i.e. averaged value of the absorption in each spectral band) or the Mean-Rosseland approximation \citep{rosseland_note_1924} are often inaccurate \citep[e.g.][]{goody_correlated-k_1989,modest_radiative_2021}. A satisfactory solution is brought by correlated-$k$ tables \citep{liou_introduction_1980, lacis_description_1991, fu_correlated_1992}. The strategy is to pre-compute high resolution absorption spectra for various pressure, temperature and mixing-ratio conditions. Then, these spectra are divided into several spectral intervals for which the opacity distribution is derived. The absorption is stored in terms of the cumulative probability in each interval \citep{fu_correlated_1992}.
It has been shown many times that this method is able to provide almost exact opacities by interpolating within the pressure, temperature and mixing-ratio grids \citep[e.g.][]{amundsen_treatment_2017, chaverot_how_2022}.
The issue is that, for each change of the atmospheric composition in the climate model, a new correlated-$k$ table needs to be created. Calculating the thousands of high resolution spectra needed for a single table is time-consuming and requires the relevant spectroscopic knowledge. The long computation time induced by large databases such as HITEMP or ExoMOL, which include many weak lines that are non negligible at high temperature, requires to run \texttt{SpeCT} on high computation facilities.
An "online" mix of correlated-$k$ tables is possible, done directly by the climate model, which allows a greater flexibility in terms of atmospheric composition \citep{amundsen_treatment_2017, dollone_modelisation_2020}. However, this method neglects the influence of the composition on the absorption line shape (e.g. the foreign collisional broadening).
It is thus only valid when one gas dominates the atmosphere. For other compositions, a corresponding correlated-$k$ table must be used, or created if it does not exist. 

Each absorption spectrum of a gas mixture containing CO$_2$ and/or H$_2$O is composed of a huge number of individual spectral lines, and each correlated-$k$ table is based on hundreds of high resolution spectra. While the core of the absorption lines is rather well represented by a Voigt profile, there is no physically based function describing the far-wings. Therefore, empirical lineshape correction factors, named $\chi$ factors, have been introduced in order to correctly model laboratory experiments. This is the most complex step in creating opacity data, but a prerequisite to guaranty accurate climate simulations.

The first aim of the present work is to provide new and updated $\chi$ factors of CO$_2$ broadened by different gases, using the latest version of the HITRAN spectroscopic database \citep{gordon_hitran2020_2021}. According to the literature \cite[e.g.,][]{forget_possible_2014,woitke_coexistence_2021}, the gases considered here (CO$_2$, H$_2$O and N$_2$) are relevant for climate modeling of terrestrial (exo)planets (e.g. temperate planets, magma ocean planets, Venus, Mars).
To do so, we reviewed and collected all the laboratory measurements of continua available in the literature for pure CO$_2$ and CO$_2$ diluted in N$_2$ and H$_2$O\footnote{\url{https://doi.org/10.5281/zenodo.15564294}}. We also homogenized the format of the existing $\chi$ factors to allow an easy implementation in other models producing opacity data for climate modeling. 
Several generic new correlated-$k$ tables dedicated to climate simulations of terrestrial planets are provided, for various gas mixtures containing H$_2$O and CO$_2$; as well as CO$_2$-CO$_2$, CO$_2$-N$_2$ and CO$_2$-H$_2$O continua\footnote{\url{https://doi.org/10.5281/zenodo.15564548}}. Finally, we propose an open-source Fortran code, named \texttt{SpeCT}\footnote{\url{https://gitlab.com/ChaverotG/spect-public}}, optimized in order to efficiently calculate high resolution spectra for various gas mixtures, for total pressures from a few Pa to hundreds of bars, and temperatures from tens to thousands of Kelvin. This code has been used to produce the tables attached to this article.

In the following sections, we first give the equations used to calculate absorption spectra in \sect{sec:stateoftheart}, as well as existing collision induced absorption an continua (\sect{sec:CIA}).
In a second step, we present \texttt{SpeCT} and its specificities (\sect{sec:spect}). Finally, in \sect{sec:results} we present the updated values of the $\chi$ factors.

\section{Computing the CO$_2$ absorption coefficient}\label{sec:stateoftheart}

The following subsections give the equations used to compute absorption spectra in the IR and in the visible \citep{hartmann_collisional_2021}.
Our approach is based on the HITRAN database \citep{rothman_hitran_2005, rothman_hitran2012_2013, gordon_hitran2016_2017, gordon_hitran2020_2021}, that is one of the most used databases in planetary science. Note that it is still valid for other spectroscopic databases such as HITEMP \citep{rothman_hitemp_2010}, ExoMol \citep{chubb_exomolop_2021} or the NASA AMES linelist \citep[e.g., for CO$_2$:][]{huang_ai-3000k_2023}. However, the $\chi$ factors are empirical corrections that are linelist dependent. This point is discussed in \sect{sec:discussion}.

\subsection{Definition of the absorption line profile}
A spectrum is made of a collection of individual optical transitions, induced by absorption/emission of photons at various wavelengths.
Unfortunately, there is, so far,  no physically based and self-consistent equation able to describe the shape of an absorption line from the center to the far-wings. The line shape near the center follows a Voigt profile (\sect{sec:line-center}) while, further in the wings, corrections factors are required to match experiment data (\sect{sec:far-wings}).

\subsubsection{Intensity of the absorption line}

The integrated intensity of each absorption - or emission - line l ($S^*_{\rm l}$ in cm$^{-1}/(\rm molecule.cm^{-2}$)) corresponding to a transition between energy levels $i$ and $j$ is given by \eq{eq:S} \citep{simeckova_einstein_2006}. 

\begin{equation}\label{eq:S}
    S^*_{\rm l}(T)=I_{\rm a}\frac{A_{\rm ji}}{8\pi c\sigma^2_{\rm l}} \frac{g_{\rm j}e^{-hcE_{\rm i}/k_{\rm B}T}\left(1-e^{-hc\sigma_{\rm l}/k_{\rm B}T}\right)}{Q(T)}
\end{equation}
where $I_{\rm a}$ is the natural terrestrial isotopic abundance  of the molecule (as given in the HITRAN database), $A_{\rm ji}$ is the Einstein coefficient (coefficient of spontaneous emission, in s$^{-1}$), $E_{\rm i}$ the lower-state energy of the transition (cm$^{-1}$), $g_{\rm j}$ the upper level statistical weight\footnote{$g_{\rm j}$ is the number of quantum states of energy $E_{\rm j}$} and $\sigma_{\rm l}$ the wavenumber of the transition (cm$^{-1}$) in vacuum. 
$Q(T)$ is the total internal partition sum over all the allowed energy levels:
\begin{equation}\label{eq:Q}
    Q(T)=\sum_{\rm k} g_{\rm k} e^{-hcE_{\rm k}/k_{\rm B}T}
\end{equation}
The value of $S^*_{\rm l}(T)$ (\eq{eq:S}) is directly given by HITRAN and HITEMP at a reference temperature $T_{\rm ref}=296$~K.
From this value, it is possible to compute the intensity at any temperature $T$:
\begin{equation}\label{eq:S(T)}
    S_{\rm l}(T)=S^*_{\rm l}(T_{\rm ref})\frac{Q(T_{\rm ref})}{Q(T)} \frac{1-e^{\frac{-hc\sigma_{\rm l}}{k_{\rm B}T}}}{1-e^{\frac{-hc\sigma_{\rm l}}{k_{\rm B}T_{\rm ref}}}} e^{\frac{-hcE_i}{k_{\rm B}}\left(\frac{1}{T}-\frac{1}{T_{\rm ref}} \right)}
\end{equation}

All the parameters of these equations are provided by the HITRAN and HITEMP databases. Note that many hot lines (with larges values of $E_{\rm i}$) are absent from HITRAN \citep{rothman_hitran_2005, rothman_hitran2012_2013, gordon_hitran2016_2017, gordon_hitran2020_2021} because they lead to negligible absorption under the temperature conditions of the Earth atmosphere. When high temperatures are involved (typically higher than 400~K), it is thus necessary to use a more relevant database, such has HITEMP \citep{rothman_hitemp_2010,hargreaves_updating_2024}.

Equation~\ref{eq:S(T)} shows that the only thermodynamic parameter influencing the intensity of an absorption line is the temperature. However, the spectral shapes of the lines are affected by different mechanisms. 
First, due to the Heisenberg uncertainty principle, there is a natural width that is inversely proportional to the lifetime of the involved j excited level. 
Second, when a molecule is not isolated, collisions happen in the gas and the lifetime of the coherence of the rotating molecular dipole is reduced. It is generally much shorter than the timescale of spontaneous emissions, and is inversely proportional to the pressure. This induces a pressure broadening of the absorption lines which is generally considerably larger than the natural broadening.
Third, the velocity of the molecules follows a Maxwell-Boltzmann distribution (function of the temperature). This induces a broadening of the absorption lines through the Doppler effect: the Doppler broadening.

In the following sections, we present the equations used to calculate the absorption spectrum of the species X interacting with N-1 other types of molecules. Therefore, the collisional parameters related to an interaction between the species X and a species y are indexed X-y.

\subsubsection{Line shape close to the center}\label{sec:line-center}

The shape of an absorption line depends on the temperature $T$, the total pressure $P$, and the gas composition, and follows a function $f(\sigma, P,T,x)$, where $\sigma$ is the current wavenumber (in cm$^{-1}$), and $x$ the set of volume mixing ratios of the considered species.
For an absorption line l centered at a wavenumber $\sigma_{\rm l}$, the influence of the pressure broadening is usually described by a Lorentz profile:
\begin{equation}
    f_{\rm Lorentz, l}(\sigma,P,T,x) = \frac{1}{\pi}\frac{\Gamma_{\rm l}(P,T,x)}{\Gamma_{\rm l}^2(P,T,x)+[\sigma-(\sigma_{\rm l} + \Delta_{\rm l}(P,T,x))]^2}
\end{equation}
The pressure has an effect on the position of each individual line which is taken into account through the pressure-induced shift $\Delta(P,T,x)$ (in cm$^{-1}$), defined as follows, for a line of a species X interacting with $N$ molecular species (including X):
\begin{equation} \label{eq:delta}
    \Delta_{\rm l}(P,T,x) = \sum_{\rm y=1}^N x_{\rm y}\frac{P}{P_{\rm ref}}\left[\delta_{\rm X-y ,l}(P_{\rm ref},T_{\rm ref}) + \delta'_{\rm X-y ,l}(T-T_{\rm ref})\right]
\end{equation}
where $\delta_{\rm X-y ,l}(P_{\rm ref},T_{\rm ref})$ are the pressure shifting coefficient (in cm$^{-1}$ atm$^{-1}$) at reference pressure and temperature of the line l of species X induced by collisions with species y, $x_{\rm y}$ is the volume mixing ratio of the species y and $\delta'_{\rm X-y ,l}$ are the temperature dependence parameters of the pressure shifting coefficient.

The half-width at half-maximum (HWHM) $\Gamma_{\rm l}(P,T,x)$ (in cm$^{-1}$) is computed using the following equation:
\begin{equation} \label{eq:gamma}
    \Gamma_{\rm l}(P,T,x)= \sum_{\rm y=1}^N x_{\rm y}\frac{P}{P_{\rm ref}}\gamma_{\rm X-y ,l}(P_{\rm ref},T_{\rm ref}) \left( \frac{T_{\rm ref}}{T}\right)^{n_{\rm X-y ,l}} 
\end{equation}
where $\gamma_{\rm X-y ,l}(P_{\rm ref},T_{\rm ref})$ is the pressure broadening coefficient (in cm$^{-1}$.atm$^{-1}$) at reference pressure and temperature of the line l of species X induced by collisions with species y and $n_{\rm X-y ,l}$ is the temperature dependence exponents of this pressure broadening. Unfortunately, some aforementioned parameters are sometimes missing in the databases, as discussed in \sect{sec:spect}. 

As molecules are not static in a gas, the absorption lines are broadened by the Doppler effect, described by a Gaussian profile:
\begin{equation}
    f_{\rm Gauss,l}(\sigma,T) = \sqrt{\frac{\ln(2)}{\pi\Gamma^2_{\rm D,l}}} e^{-\frac{\ln(2)(\sigma-\sigma_{\rm l})^2}{\Gamma^2_{\rm D,l}}}
\end{equation}
where $\Gamma_{\rm D}(T)$ is the Doppler HWHM defined as:
\begin{equation}
    \Gamma_{\rm D,l}(T) = \frac{\sigma_{\rm l}}{c} \sqrt{\frac{2{\rm N_{\rm a}}k_{\rm B}T\ln(2)}{M}} 
\end{equation}

with $M$ the molar mass of the absorber and $\rm N_{\rm a}$ the Avogadro number.
The most usual way to account for both pressure and Doppler broadening is to use the Voigt profile, which is a convolution of a Lorentz and a Gauss functions.
The absorption cross-section $\eta_{\rm X}(\sigma, T, P)$ (in cm$^2.\rm molecule^{-1}$) at a wavenumber $\sigma$ of a species X, is given by the sum of all individual line contributions:
\begin{equation}\label{eq:sigma}
    \eta_{\rm X}(\sigma,P,T,x) = \sum_{\rm l=1}^{\rm L} S_{\rm l}(T) \times \left[ f_{\rm Lorentz, l}(\sigma,P,T,x)\circledast f_{\rm Gauss, l}(\sigma,T)\right]
\end{equation}
A convenient quantity derived from \eq{eq:sigma} is the absorption coefficient $k_{\rm X}(\sigma,P,T)$ (in cm$^{-1}$) defined as: 
\begin{equation}\label{eq:kappa}
    k_{\rm X}(\sigma,P,T,x) = N_{\rm X} \eta_{\rm X}(\sigma,P,T,x) = \frac{x_{\rm X}P}{10^6k_{\rm B}T} \eta_{\rm X}(\sigma,P,T,x)
\end{equation}
where $N_{\rm X}$ is the number of molecules of species $X$ per unit volume (in $\rm molecules. \rm cm^{-3}$). This quantity is expressed in \eq{eq:kappa} as a function of the pressure and the temperature by using the ideal gas law, with $x_{\rm X}P$ (in Pa) the partial pressure of the considered absorbing species.

Farther in the wings, the absorption significantly deviates from the Voigt profile and corrections factors are required to accurately calculate the line shape as discussed below.

\subsection{Absorption in the far-wings: the necessity of $\chi$ factors}\label{sec:far-wings}

\begin{figure*}[!ht]
    \centering\includegraphics[width=\linewidth]{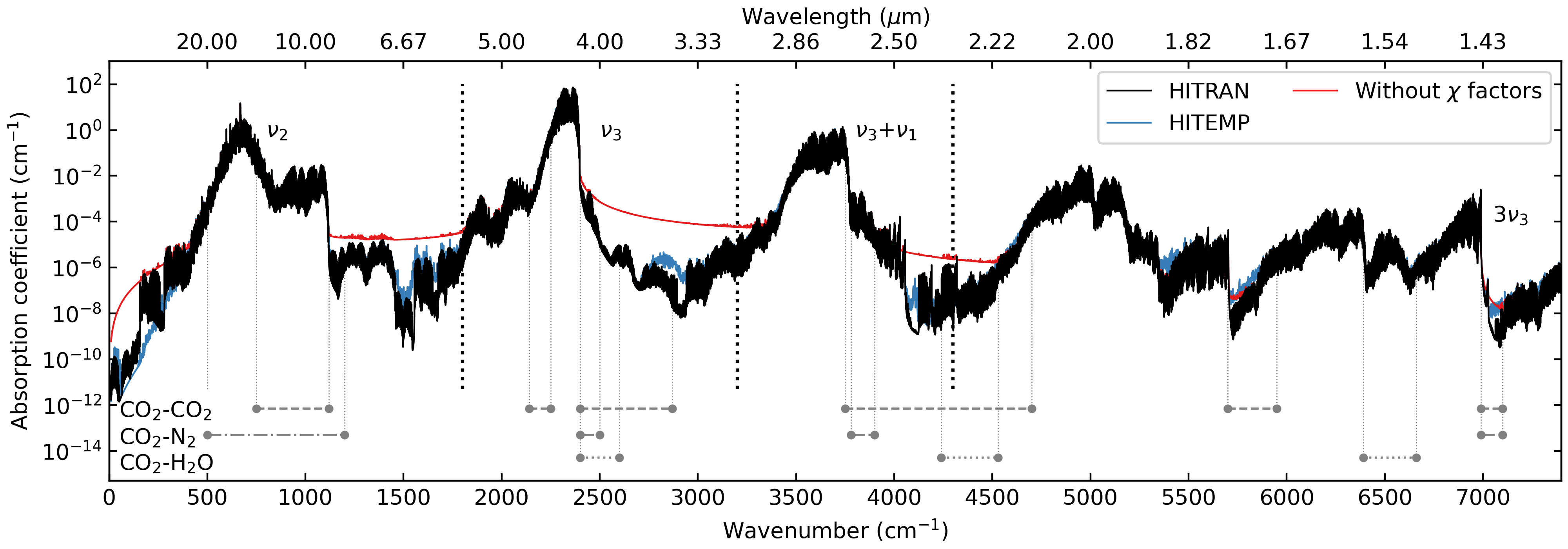}
    \caption{Comparison of calculated spectra for 1~bar of pure CO$_2$ gas at 750~K, with (black line) and without (red line) the $\chi$ factors. The blue line is the spectrum calculated for the same pressure and temperature conditions, but using HITEMP 2024 \citep{hargreaves_updating_2024} instead of HITRAN 2020 \citep{gordon_hitran2020_2021} The vertical dotted black lines illustrate the definition interval of the 3 main bands of CO$_2$ we consider. The horizontal dashed, dot-dashed and dotted gray lines represent the intervals for which measurement of the continuum (excluding CIAs) are available, for CO$_2$-CO$_2$, CO$_2$-N$_2$ and CO$_2$-H$_2$O, respectively.}
    \label{fig:bands_exp}
\end{figure*}

There is no physically based function able to accurately model the line shape from the center of the line to the far-wings. Therefore, empirical correction factors, named $\chi$ factors, and usually based on laboratory measurements, are generally used.
They are functions of the distance to the line-center, of the temperature and of the absorbing and foreign species. 
Close to the core of the line, where measurements are usually well described by a Voigt profile, $\chi$=1. In the far-wings, where the Voigt profile is no more accurate, $\chi>1$ values define "super-Lorentzian" profiles, while $\chi<1$ values define "sub-Lorentzian" profiles. For instance, for a pure CO$_2$ gas, $\chi\neq1$ beyond 3~cm$^{-1}$ from the center of the absorption line \citep[e.g.,][]{perrin_temperature-dependent_1989, tran_measurements_2011}.
Without this correction step, the absorption is largely overestimated in between the CO$_2$ absorption bands, as shown by the differences between the red and black curves in \fig{fig:bands_exp}.

A classical definition of the absorption coefficient $k_X(\sigma, T, P,x)$, including $\chi$ factors, is given by \eq{eq:eq_chi_factors}. It is the sum of $L$ individual lines of a species X interacting with $N$ molecular species (including X).
\begin{equation}
\begin{split}
    k_{\rm X}(\sigma, T, P,x) = \frac{x_{\rm X}P}{10^6k_{\rm B}T} \sum_{\rm l=1}^{\rm L}   S_{\rm l}(T) \frac{hc(\sigma-\sigma_{\rm l})}{k_{\rm B}T} \frac{1}{1-\exp\left(\frac{-hc(\sigma-\sigma_{\rm l})}{k_{\rm B}T}\right)} \\ \times
     \frac{\sigma}{\sigma_{\rm l}} \frac{1-\exp\left(-\frac{hc\sigma}{k_{\rm B}T}\right)}{1-\exp\left(-\frac{hc\sigma_{\rm l}}{k_{\rm B}T}\right)}  
    \times \frac{1}{\pi}\frac{\Gamma_{\rm l}(P,T,x)}{[\sigma-(\sigma_{\rm l}+\Delta_{\rm l}(P,T,x))]^2+[\Gamma_{\rm l}(P,T,x)]^2} \\ \times
    \frac{\sum_{\rm y=1}^N x_{\rm y}\gamma_{\rm X-y ,l}(T) \chi_{\rm X-y }(T, |\sigma-\sigma_{\rm l}|)}{\Gamma_{\rm l}(P,T,x)}
\end{split} \label{eq:eq_chi_factors}
\end{equation}
where $S_{\rm l}(T)$ is the integrated line intensity (in cm$^{-1}/(\rm molecule.cm^{-2})$) from \eq{eq:S(T)}, $\Delta_{\rm l}(P,T,x)$ the pressure shift from \eq{eq:delta} and  $\Gamma_{\rm l}(P,T,x)$ the pressure broadened half width at half maximum from \eq{eq:gamma}.
Note that, if the pressure and temperature conditions are such that deviations with respect to the ideal gas law are significant (e.g., at elevated pressure), the factor $(x_{\rm X}P)/(10^6k_{\rm B}T)$ must be replaced by the true density (in molecules.cm$^{-3}$) of species X, obtained, for instance from an equation of state.
The $\frac{hc(\sigma-\sigma_{\rm l})}{k_{\rm B}T}\times\left[1-\exp\left(\frac{-hc(\sigma-\sigma_{\rm l})}{k_{\rm B}T}\right)\right]^{-1}$ term is the quantum asymmetry factor \citep{frommhold_collision-induced_1994,fakhardji_direct_2022}. 
This factor accounts for an asymmetry existing between the left and right far-wings (i.e., toward low or high wavenumbers) which is negligible close to the line-center when $|\sigma-\sigma_{\rm l}|$ tends to zero. For this reason, the asymmetry factor does not appear in \eq{eq:kappa}, used to calculate the line shape close to the line center. As shown by \cite{fakhardji_direct_2022}, the asymmetry factor used in \cite{tran_measurements_2018}, and equal to $\exp\left(\frac{hc(\sigma-\sigma_{\rm l})}{2k_{\rm B}T}\right)$, is not accurate for temperatures lower than 100~K. 

Laboratory measurements are usually available in limited portions of the spectrum.
The three most intense IR absorption bands in the CO$_2$ spectrum, are the $\nu_2$, $\nu_3$ and $\nu_3$+$\nu_1$ bands, centered at 667~cm$^{-1}$ (14.9~$\mu$m), 2349~cm$^{-1}$ (4.3~$\mu$m) and 3737~cm$^{-1}$ (2.7~$\mu$m), respectively (see \fig{fig:bands_exp}). Usually, measurements for the determination of $\chi$ factors are performed in the wings of these bands where the absorption of the continuum is intense enough to be detected. Therefore, different $\chi$ factors correct the far-wings of the three main bands of CO$_2$. 
We apply the $\chi$ factors determined for the $\nu_2$ band between 0 and 1800~cm$^{-1}$, for the  $\nu_3$ band between 1800 and 3200~cm$^{-1}$ and for the $\nu_3$+$\nu_1$ band between 3200 and 4300~cm$^{-1}$ (vertical dotted lines in \fig{fig:bands_exp}). 
For absorption lines beyond 4300~cm$^{-1}$, the $\chi$ factors of the $\nu_3$ band are, due to lack of data, systematically applied to account for the overtone bands (e.g. the 3$\nu_3$ band at 7000~cm$^{-1}$). 

To define the $\chi$ factors themselves, we adopt the formalism proposed by \cite{perrin_temperature-dependent_1989} (see also \citealt{tran_measurements_2011, tran_measurements_2018}), as presented in \tab{tab:exp_chi_factors}. The analytical expression depends on the distance to the line-center in order to account for variations of the deviation of the actual line shape from the Voigt profile. The boundaries ($\sigma_1$, $\sigma_2$ and $\sigma_3$) depend on the gas mixture and on the considered spectral band.
The $B_{\rm y}$ coefficients (in cm) in \tab{tab:exp_chi_factors} are defined as follows \citep{perrin_temperature-dependent_1989}:
\begin{equation}\label{eq:B}
    B_{\rm i}(T) = \alpha_{\rm i} + \beta_{\rm i} \exp(-\gamma_{\rm i}T)
\end{equation}
and depend on $\alpha$, $\beta$ and $\gamma$, that are functions of the collision partner y and of the spectral band. These coefficients and the boundaries ($\sigma_1$, $\sigma_2$ and $\sigma_3$) are given in \sect{sec:results} (\tabs{tab:new_chi_factors}{tab:interval_chi_factors}).

\begin{table*}[ht!]
	\centering
    \caption{Analytical expression of $\chi$ factors, as a function of the distance to the line-center ($\Delta\sigma$).}
	\begin{tabular}{ l l}
        \hline  \hline  
        $\chi(T, \Delta\sigma)$  & $\Delta\sigma$   \\
        \hline 
        1 & $0<\Delta\sigma<\sigma_1$  \\
        $\exp[-B_1(|\Delta\sigma|-\sigma_1)]$  &  $\sigma_1<\Delta\sigma<\sigma_2$ \\
        $\exp[- B_1(\sigma_2-\sigma_1)-B_2(|\Delta\sigma|-\sigma_2)]$ & $\sigma_2<\Delta\sigma<\sigma_3$ \\
        $\exp[- B_1(\sigma_2-\sigma_1)-B_2(\sigma_3-\sigma_2)-B_3(|\Delta\sigma|-\sigma_3)]$ & $\sigma_3<\Delta\sigma$ \\
        \hline 
	\end{tabular}

 \label{tab:exp_chi_factors}
\end{table*}

\section{Existing Collision Induced Absorption (CIA) and continua}\label{sec:CIA}

The equations given above only describe the contributions to the absorption coefficient due to the intrinsic dipole moment (induced by vibration in the case of CO$_2$) of species X, but when two molecules collide, this can 1) briefly induce a transient dipole through, for instance the polarization of CO$_2$ by the electric field of the multipoles of the collision partner \citep[e.g.][]{frommhold_collision-induced_1994,karman_o2o2_2018}, 2) form a pair, leading to the creation of a bound molecular complex currently denoted as dimer.
These absorption sources must be taken into account when calculating the radiative transfer in an atmosphere. For this reason, we compiled and homogenized all the relevant CIAs, dimers and continua existing in the literature for the gas mixture we consider. These data are described below and available at: \href{https://doi.org/10.5281/zenodo.15564548}{https://doi.org/10.5281/zenodo.15564548}.

Several studies have proposed measurements or calculations of the CIA and data are available through the HITRAN CIA\footnote{\url{https://hitran.org/cia/}} database \citep{karman_update_2019}. 
These two processes, CIA and dimers, are the only significant absorption source of symmetric molecules such as N$_2$ or H$_2$. 
For radiatively active molecules such as H$_2$O and CO$_2$, CIA and dimers can generate absorption in between the bands due to the intrinsic dipole (e.g., the $\nu_1$ and 2$\nu_2$ CIA bands of CO$_2$ between 1200 and 1500 cm$^{-1}$ in Fig. 1), modifying the radiative effect of the gases. 
To follow the convention of climate models, we do not include the CIAs directly in the correlated-$k$, but we account for them in the continua files.

For CO$_2$-CO$_2$, the CIA band at 0-250~cm$^{-1}$ \citep{gruszka_roto-translational_1997}, is extrapolated down to 100~K. We added every CIA and dimer bands listed in \cite{tran_collision-induced_2024}, using the associated proposed functional formula for the shape and band intensity (as a function of temperature). The temperature dependence extends from 100~K to 800~K.

For the N$_2$-N$_2$ CIA, we combined all data (roto-translational, fundamental and first overtone bands) from \cite{karman_update_2019}\footnote{available at: \url{https://hitran.org/cia/}}. The temperature dependence is extrapolated from 70 to 500~K. 

The continua of H$_2$O-H$_2$O and H$_2$O-N$_2$ are provided by MT\_CKD \citep{mlawer_inclusion_2023}. 
However, for the second mixture, H$_2$O is broadened by "air", that corresponds to the present-day Earth's atmosphere ratio of O$_2$ and N$_2$. We included in these continua the N$_2$-H$_2$O CIAs from \cite{hartmann_collision-induced_2017} and \cite{baranov_h2on2_2012}.

For H$_2$O-CO$_2$, the continuum due to the wings of H$_2$O lines broadened by collisions with CO$_2$ has been recalculated following the procedure described in \cite{tran_measurements_2018} (using HITRAN 2016 line list complemented with CO$_2$-broadening coefficients) and extended up to 20000~cm$^{-1}$. The temperature dependence was calculated using \cite{ma_far_1992}, up to 10000~cm$^{-1}$ (no data beyond).
We also added the simultaneous H$_2$O+CO$_2$ CIA band, recently measured near 6000~cm$^{-1}$ from \cite{fleurbaey_simultaneous_2022}. Note that the temperature dependence is not known for this CIA, thus we assume that there is not temperature dependency.

All the CIAs and continua joined to this article are given in cm$^{-1}$.amagat$^{-2}$. Note that 1~amagat corresponds to the density at standard temperature/pressure ($P_0$=1 atm and $T_0$=273.15~K) which is given by the Loschmidt constant=2.686$\times10^{25}$~molecules/m$^3$. For an ideal gas at temperature T and pressure P, the corresponding density in amagat units is simply $(P\times T_0)/(P_0\times T)$. Therefore, the absorption coefficient $k_{\rm X-y }(\sigma, T)$ (in cm$^{-1}$) of a species X colliding with a species y is given by:
\begin{equation}
    k_{\rm X-y }(\sigma, T, P, x) = A_{\rm X-y}(\sigma, T)x_{\rm X}x_{\rm y} \left[\frac{273.15\times P}{101325\times T}\right]^2
\end{equation}
where $A_{\rm XY}(\sigma, T)$ is the absorption coefficient in cm$^{-1}$.amagat$^{-2}$, $x_{\rm X}$ and $x_{\rm Y}$ are the mixing ratio of the different species, $P$ the pressure (in Pa) and $T$ is the temperature (in K). If the pressure and temperature conditions are such that the ideal gas law does is not valid, the squared term must be replaced by the real squared density calculated using an equation of state.
Outside the given temperature range, we advise keeping the CIAs constant at the closest known temperature (i.e. no extrapolation).

\section{Technical description of \texttt{SpeCT} } \label{sec:spect}

The code\footnote{\url{https://gitlab.com/ChaverotG/spect-public}} is written in Fortran and is highly parallelized using MPI and openMP in order to efficiently compute the hundreds of spectra required to produce a correlated-$k$ table. 
The calculations of the core of the lines (absorption up to $\pm$25~cm$^{-1}$ from the line-center) and of the far-wings (beyond $\pm$25~cm$^{-1}$) are done separately, following the historical convention proposed by the MT\_CKD consortium \citep{mlawer_development_2012, mlawer_analysis_2019}.
For computation time reasons, we apply a cutoff to the line wings. This avoids the calculation of the parts of the far-wings which do not contribute to the total absorption spectrum.
Sensitivity tests done on this value showed that $\pm$1500~cm$^{-1}$ from the line-center is required to accurately compute the continua. More precisely, a cutoff beyond 1500~cm$^{-1}$ does not affect the value of the continuum. A cutoff closer to the line-center lowers the continuum, especially in weak absorption regions. As the aim is to produce correlated-$k$ tables and accurate continua, we advise keeping this value unchanged. This cutoff value implies that most of the time, the $\chi$ factors are extrapolated outside the interval in which they have been calculated.

To summarize, \texttt{SpeCT} computes: 1) high resolution spectra containing the centers of the lines from 0 to $\pm$25~cm$^{-1}$ using a Voigt profile (when $\chi=1$) or a corrected Lorentzian profile (\eq{eq:eq_chi_factors}, when $\chi\neq1$), 2) the part of the spectrum containing the far-wings (from $\pm$25~cm$^{-1}$ to $\pm$1500~cm$^{-1}$) using \eq{eq:eq_chi_factors}. Note that in the wings (i.e. far from the line center) the Lorentzian and Voigt profiles are equivalent.

As the continua (made of the sum of the far-wings) normalized by $P^2$ do not depend on the pressure, they are computed only for a range of temperature (given in the Zenodo repository\footnote{\url{https://doi.org/10.5281/zenodo.15564548}}). Also, the slow variation of the continua with the wavenumber allows computing them at a lower resolution, in order to reduce the computation time.
The center of the lines are computed at a resolution equal to 1$\times10^{-3}$~cm$^{-1}$, that is a good compromise between accuracy and computation time, while the continua are computed with a step of 5~cm$^{-1}$.

In both case, the structure is the following: 1) reading of the spectroscopic linelists (HITRAN or HITEMP), 2) distribution of the list of initial conditions (temperature, pressure and mixing-ratios) across MPI processes, 3) computation of the profile of each individual line using equations given in \sect{sec:stateoftheart}, 4) sum of individual profiles in order to obtain a final spectrum.

The continua for H$_2$O-H$_2$O and H$_2$O-N$_2$ are taken from MT\_CKD v\_4.0.1 \citep{mlawer_inclusion_2023}\footnote{\url{https://github.com/AER-RC/MT_CKD_H2O}} and the H$_2$O-CO$_2$ continuum has been recalculated as described in \sect{sec:CIA}.
The continua of CO$_2$ are computed using \eq{eq:eq_chi_factors}, and the $\chi$ factors proposed in this work. 
Following \citet{mlawer_development_2012} we include in the continuum the plinth of the absorption lines, that is the pedestal of the line-center assuming a constant value equal to the absorption at $\pm$25~cm$^{-1}$ from the line-center. To be consistent, we remove the plinth from the calculation of the core of the lines. 
We do not include the CIAs in the absorption spectra used to create the correlated-$k$ tables, and we give separately the CIAs (provided by previous articles, see \sect{sec:CIA}) and the CO$_2$ continua calculated in this work.  

\texttt{SpeCT} gives as outputs one spectrum per pressure/temperature/mixing ratios condition, in a simple precision binary format, as a function of the wavenumber. The wavenumber grid is given in a separate file. This allows to minimize the size of the outputs, allowing the storage of thousands of spectra. The spectra corresponding to the line-centers regions are outputted in cm$^{-1}$ while the continua are given in cm$^{-1}$.amagat$^{-2}$.
Thanks to a joint development, spectra calculated by \texttt{SpeCT} can be read automatically by \texttt{Exo\_k}\footnote{\url{https://perso.astrophy.u-bordeaux.fr/~jleconte/exo_k-doc/index.html}} \citep{leconte_spectral_2021} to create correlated-$k$ tables.

As mentioned in the \sect{sec:stateoftheart}, many hot lines are absent from HITRAN \citep{gordon_hitran2020_2021}. To circumvent this issue, we use HITEMP 2010 for H$_2$O \citep{rothman_hitemp_2010}, or HITEMP 2024 for CO$_2$ \citep{hargreaves_updating_2024}, to compute spectra beyond 400~K. This means that the correlated-$k$ we provide are hybrid and based on HITRAN and HITEMP to guaranty a great accuracy at both low and high temperatures.

Unfortunately, all the line parameters, described in \sect{sec:stateoftheart}, are not available in HITRAN yet (even if known for some of them).
In both databases, the temperature dependence coefficients of the pressure shift ($\delta'_{\rm X-y ,l}$ in \ref{eq:delta}) are not included for both H$_2$O and CO$_2$ absorption lines, meaning that in \texttt{SpeCT} we assume $\delta'_{\rm X-y ,l}=0$ for H$_2$O-H$_2$O, H$_2$O-CO$_2$, CO$_2$-CO$_2$ and CO$_2$-H$_2$O.
In HITRAN, for CO$_2$ absorption lines, the foreign pressure shift coefficient $\delta_{\rm CO_2-H_2O}$ at the temperature $T_{\rm ref}$ is missing.
For H$_2$O absorption lines, the only pressure shift coefficient available is the $\delta_{\rm H_2O-air}$ (both $\delta_{\rm H_2O-H_2O}$ and $\delta_{\rm H_2O-CO_2}$ are missing). As suggested by \cite{brown_co2-broadened_2007}, we calculate $\delta_{\rm H_2O-CO_2}$ as the air coefficient $\delta_{\rm H_2O-air}$ multiplied by 1.67.
Also, the pressure broadening coefficient $\gamma_{\rm H_2O-CO_2}$ is missing, as well as the temperature dependence $n_{\rm H_2O-H_2O}$ and $n_{\rm H_2O-CO_2}$. Therefore, the HWHM (\eq{eq:gamma}) of H$_2$O lines broadened by CO$_2$ and N$_2$ becomes:
\begin{equation}
\begin{split}
    \Gamma_{\rm H_2O,l}(P,T,x)= \left( \frac{T_{\rm ref}}{T}\right)^{n_{\rm H_2O-air,l}} \\
    \times P\left[x_{\rm H_2O} \gamma_{\rm H_2O-H_2O,l}(P_{\rm ref},T_{\rm ref})
    + (1-x_{\rm H_2O}) \gamma_{\rm H_2O-air,l}(P_{\rm ref},T_{\rm ref}) \right] 
\end{split}
\end{equation}
Even if not included in HITRAN yet, some of these parameters exist in the literature, thus we plan to add them in a future development of \texttt{SpeCT}.
In HITEMP, for both H$_2$O and CO$_2$, the only available parameters are $\gamma_{\rm X-X,l}(P_{\rm ref},T_{\rm ref})$, $\gamma_{\rm X-air,l}(P_{\rm ref},T_{\rm ref})$, $n_{\rm X-air,l}$ and $\delta_{\rm X-air}$.

\section{Results}\label{sec:results}

\subsection{Revisited $\chi$ factors}\label{sec:new_chi_factors}

Based on the formalism presented in \sect{sec:stateoftheart}, we adjusted the $\chi$ factors using different laboratory measurements available in the literature. A complete list of the existing experimental data is given in \app{app:refs}. We also provide the laboratory measurements we used in an ascii format at: \href{https://doi.org/10.5281/zenodo.15564294}{https://doi.org/10.5281/zenodo.15564294}. 
This bibliographic work could be useful for further adjustment of the $\chi$ factors, following improvements of the linelists. 
The $\chi$ factors provided in this work are given in \tab{tab:new_chi_factors} while the corresponding cutoff distances are in \tab{tab:interval_chi_factors}. 

Even if some $\chi$ factors have been derived recently with modern versions of HITRAN (e.g., $\nu_2$ band of pure CO$_2$ from \citealt{tran_measurements_2011} and $\nu_3$ band of CO$_2$-H$_2$O from \citealt{tran_measurements_2018}), we chose to re-adjust them anyway to guaranty a correct match with the change of asymmetry factor (see \sect{sec:far-wings}).

\begin{figure}[!ht]
    \centering\includegraphics[width=\linewidth]{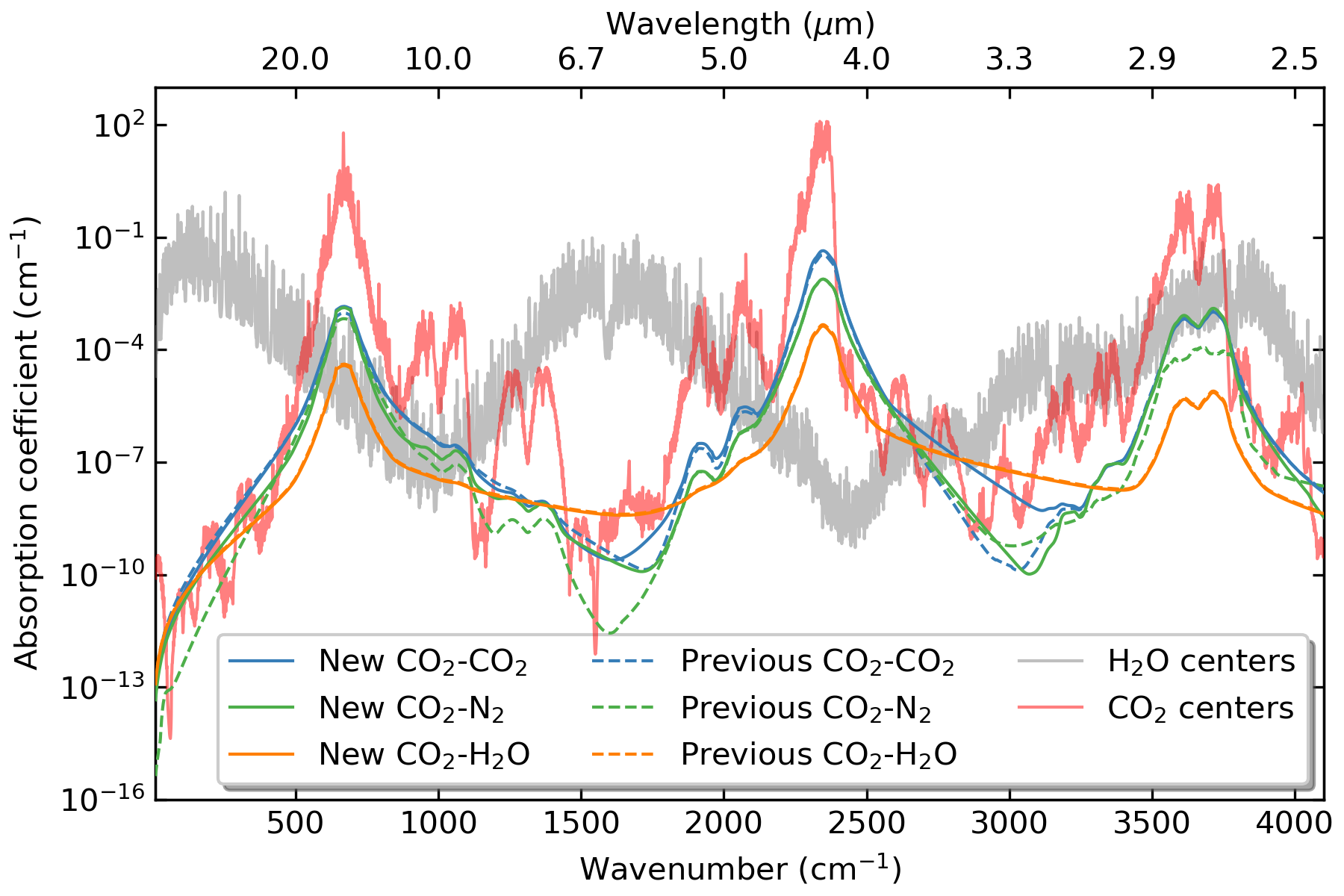}
    \caption{Absorption spectrum for a gas mixture including 0.01~bar of H$_2$O, 1~bar of CO$_2$ and 1~bar of N$_2$ at 300~K. Red and grey lines are the spectra containing the centers of absorption lines (of CO$_2$ and H$_2$O respectively) up to $\pm$25~cm$^{-1}$. Other solid lines are the CO$_2$ continua obtained using the original $\chi$ factors from this work, while the dashed lines are the spectra calculated with previously existing factors. 
    The continua calculated with existing $\chi$ factors (dashed lines) are based on \cite{tran_measurements_2011}, \cite{perrin_temperature-dependent_1989} and \cite{tran_measurements_2018} for pure CO$_2$, CO$_2$-N$_2$ and CO$_2$-H$_2$O, respectively. For this comparison, we do not use $\chi$ factors from \cite{burch_absorption_1969} to calculate the $\nu_1+\nu_3$ band of CO$_2$-N$_2$ as it is based on a different formalism, thus the factors derived for the $\nu_3$ band \citep{perrin_temperature-dependent_1989} are applied everywhere (green dashed line).}
    \label{fig:new_chi}
\end{figure}

An overview of the changes induced by the new $\chi$ factors is given by \fig{fig:new_chi}, for a gas mixture including 0.01~bar of H$_2$O, 1~bar of CO$_2$ and 1~bar of N$_2$ at 300~K. The different continua obtained using our correction are given by colored lines, while the ones calculated using $\chi$ factors from the literature (see references below) are given by dotted colored lines. Water and CO$_2$ local lines contributions (gray and red lines, respectively) are also indicated in the figure for comparison. The next subsections describe the data and the method used to derive the new $\chi$ factors of CO$_2$ broadened by CO$_2$ (\sect{sec:CO2CO2}), N$_2$ (\sect{sec:CO2N2}), and H$_2$O (\sect{sec:CO2H2O})

\begin{table*}
	\centering
 \caption{Values of the $\chi$ factors parameters for CO$_2$ mixtures defined in \tab{tab:exp_chi_factors}. }
	\begin{tabular}{l l l l c l l l c l l l }
        \toprule
         & \multicolumn{3}{c}{$\nu_2$ band ($\approx$ 667~cm$^{-1}$)} && \multicolumn{3}{c}{$\nu_3$ band ($\approx$ 2349~cm$^{-1}$)} && \multicolumn{3}{c}{$\nu_1+\nu_3$ band ($\approx$ 3737~cm$^{-1}$)} \\
        \midrule  
        CO$_2$-CO$_2$ & $\alpha$ & $\beta$ & $\gamma$  && $\alpha$ & $\beta$ & $\gamma$ && $\alpha$ & $\beta$ & $\gamma$ \\
        \cmidrule{2-4} \cmidrule{6-8} \cmidrule{10-12}
        B$_1$ & 0.085 & 1.962 & 0.02 && 0.055 & -0.11 & 0.008 && 0.034 & -0.4 & 0.0162 \\
        B$_2$ & 0.0185 & - & - && 0.021 & 0.08 & 0.0095 && 0.016 & 2174.3 & 0.057  \\
        B$_3$ & 0.011 &-  & -  && 0.01 & - & -  && 0.01 & - & - \\
        \midrule 
        CO$_2$-N$_2$ & $\alpha$ & $\beta$ & $\gamma$  && $\alpha$ & $\beta$ & $\gamma$ && $\alpha$ & $\beta$ & $\gamma$ \\
        \cmidrule{2-4} \cmidrule{6-8} \cmidrule{10-12}
        B$_1$ & 0.065 & 0.038 & 0.003 &&  0.416 & -0.354 & 0.00386 && 0.02 & -0.354 & 0.0386 \\
        B$_2$ & 0.018 & 0.055 & 0.02 && 0.0167 & 0.0421 & 0.00248 && 0.33 & 0.0421 & 0.00248  \\
        B$_3$ & 0.0085 & -  & -  && 0.019 & - & -  && 0.019 & - & - \\
        \midrule
        CO$_2$-H$_2$O & $\alpha$ & $\beta$ & $\gamma$  && $\alpha$ & $\beta$ & $\gamma$ && $\alpha$ & $\beta$ & $\gamma$ \\
        \cmidrule{2-4} \cmidrule{6-8} \cmidrule{10-12}
        B$_1$ &  - & - & - && 0.075 & -0.02589 & 0.00344 && - & - & -\\
        B$_2$ & - & - & - && 0.024 & -0.0166 & 0.00199 && - & - & -  \\
        B$_3$ & - &-  & -  && 0.0025 & - & -  && - & - & - \\
        \bottomrule
	\end{tabular}

 \label{tab:new_chi_factors}
\end{table*}

\begin{table}
	\centering
    \caption{Interval definition of the $\chi$ factors for the different bands defined in \tab{tab:exp_chi_factors}.}
	\begin{tabular}{l l l l }
        \toprule 
        & $\sigma_1$ & $\sigma_2$ & $\sigma_3$ \\
        \midrule
        CO$_2$-CO$_2$  & & & \\
        \cmidrule{1-1}
        $\nu_2$ band & 3 & 30 & 150 \\
        $\nu_3$ band & 3 & 50 & 220 \\
        $\nu_1+\nu_3$ band & 3 & 120 & 300 \\ 
        \midrule 
        CO$_2$-N$_2$ & & & \\
        \cmidrule{1-1}
        $\nu_2$ band & 3 & 50 & 180 \\
        $\nu_3$ band & 3 & 10 & 70 \\
        $\nu_1+\nu_3$ band & 3 & 10 & 70 \\
        \midrule 
        CO$_2$-H$_2$O & & & \\
        \cmidrule{1-1}
        $\nu_3$ band & 5 & 35 & 170 \\
        \bottomrule
	\end{tabular}
 \label{tab:interval_chi_factors}
\end{table}

\subsubsection{CO$_2$-CO$_2$} \label{sec:CO2CO2}
They are several laboratory measurements for a pure CO$_2$ gas in the 3 main bands of CO$_2$ shown in \fig{fig:bands_exp}.
The $\chi$ factors of the $\nu_2$ band have been marginally adapted (including local bands), based on \cite{tran_measurements_2011}. We obtain the same order of magnitude of difference between the $\chi$ factors method and the experimental measurements in the central region of the bands as what is shown in \cite{tran_measurements_2011}. To overcome this issue, a more complex line-mixing calculation is required, as discussed in \sect{sec:discussion}. 

\begin{figure}[!ht]
    \centering\includegraphics[width=\linewidth]{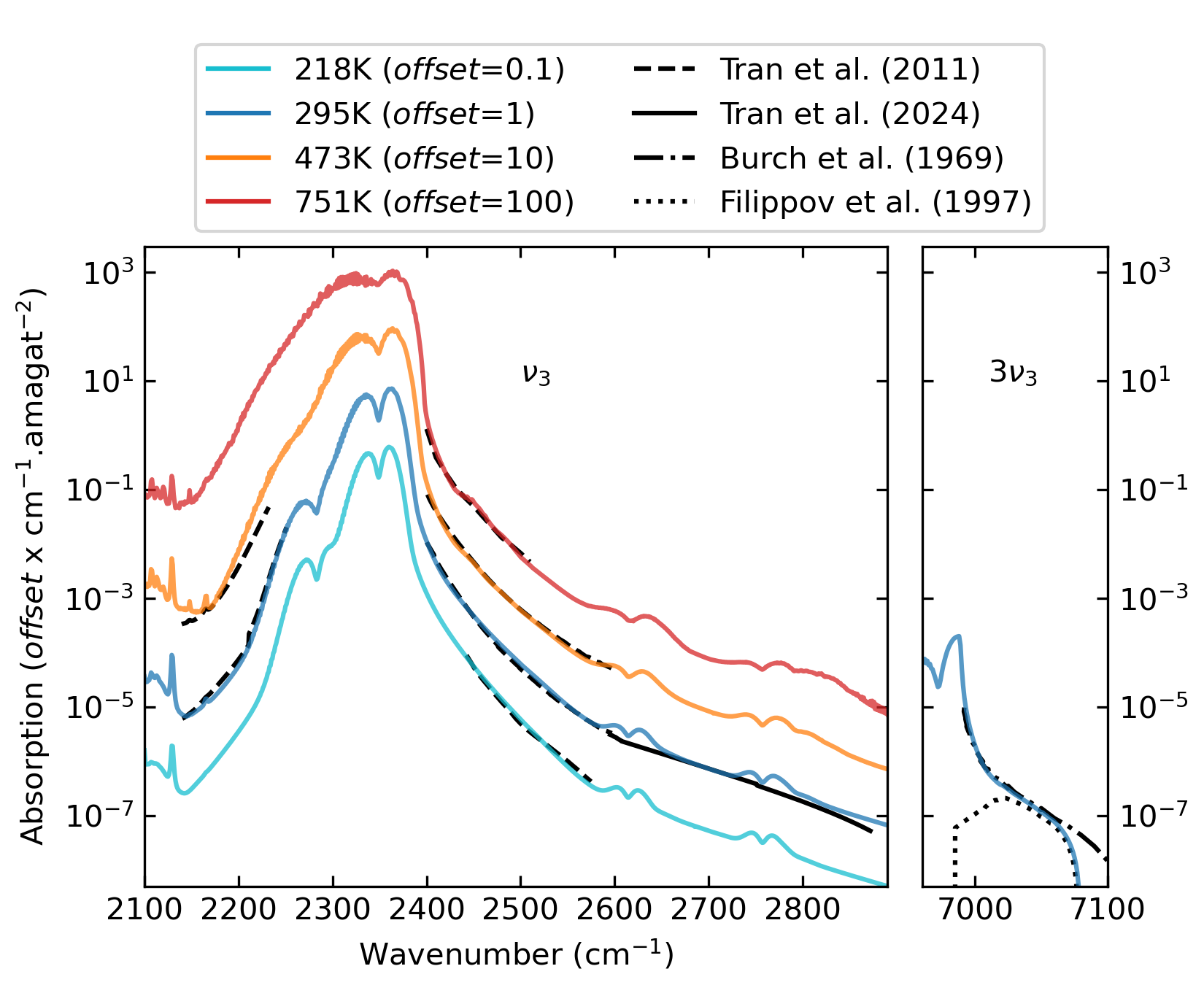}
    \caption{Adjustment of the $\chi$ factors of a pure CO$_2$ gas for various temperatures. We include the CIA, near 7050~cm$^{-1}$, from \cite{filippov_collision-induced_1997} (black dotted line on the right panel) to adjust the $\chi$ factors at 295~K. Black dashed, solid and dot-dashed lines correspond to experiment data from \cite{tran_measurements_2011}, \cite{tran_collision-induced_2024} and \cite{burch_absorption_1969}, respectively. Colored lines are the spectra calculated with our model, using the original $\chi$ factors proposed in this work. For visualization reasons, we arbitrary multiplied the spectra by an offset (see \textit{offset} on the figure) to avoid potential overlapping of the curves.}
    \label{fig:adj_CO2CO2}
\end{figure}

We also adjusted the $\nu_3$ band wing using experimental data of \cite{tran_measurements_2011}, which are limited to 2600~cm$^{-1}$. From the experimental spectra of \cite{tran_measurements_2011}, new data for the $\nu_3$ band wing in the 2600-2900~cm$^{-1}$ range have been determined thanks to the accurate determination of the CIAs in this spectral region (see \citealt{tran_collision-induced_2024}). These data are thus also used in our fits of the $\chi$ factors, as shown by \fig{fig:adj_CO2CO2}. This explains the large difference in the 2600-2900~cm$^{-1}$ range between our results (blue solid line in \fig{fig:new_chi}) and those obtained with the $\chi$ factors of \cite{tran_measurements_2011} (blue dashed line in the same figure). 
The $3\nu_3$ band is corrected with the same factors as it is an overtone band of the $\nu_3$ band (right panel of \fig{fig:adj_CO2CO2}).  
The obtained spectra fit the measurements of \cite{burch_absorption_1969} in the $3\nu_3$ region by adding the CIA from \cite{filippov_collision-induced_1997}.

The $\nu_1+\nu_3$ $\chi$ factors have been slightly adapted, based on \cite{tran_measurements_2011}, and using additional data from \cite{tonkov_measurements_1996}. 

\subsubsection{CO$_2$-N$_2$} \label{sec:CO2N2}
\begin{figure}[!ht]
    \centering\includegraphics[width=\linewidth]{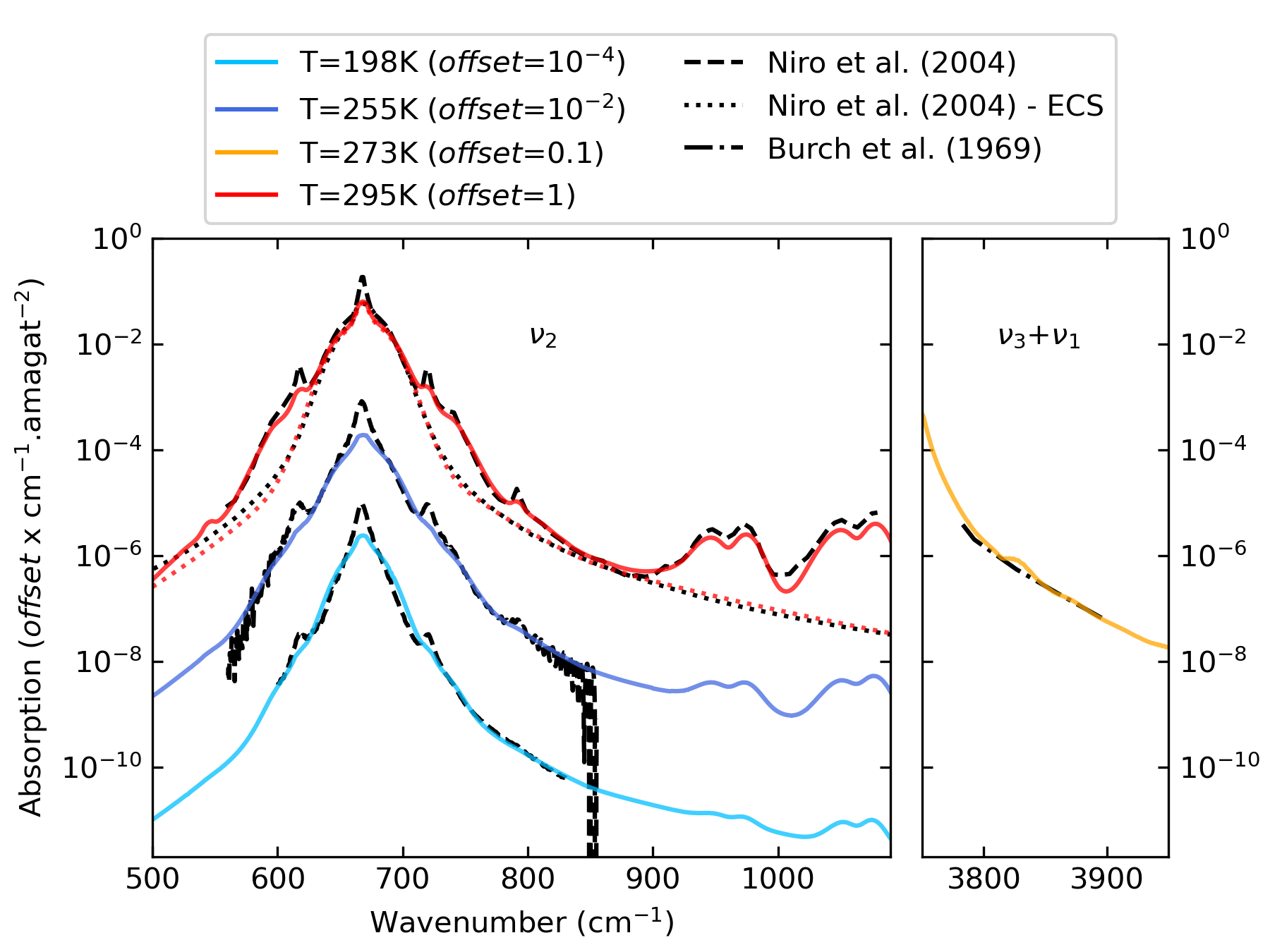}
    \caption{Adjustment of the $\chi$ factors of CO$_2$ broadened by N$_2$ for various temperatures. The black dashed and dotted-dashed lines correspond to experiment data from \cite{niro_spectra_2004} and \cite{burch_absorption_1969}, respectively. The black dotted line corresponds to a spectrum computed using the ECS model from \cite{niro_spectra_2004}, that is the core of the $\nu_2$ band without local bands. The red dotted line is our $\chi$ factor calculation of the $\nu_2$ band without local bands. For visualization reasons, we arbitrarily multiplied the spectra by a factor (see \textit{offset} on the figure) to avoid potential overlapping of the curves. }
    \label{fig:adj_CO2N2}
\end{figure}

The $\chi$ factors of the $\nu_3$ band from \cite{perrin_temperature-dependent_1989} have been marginally adapted, by using measurements of the $3\nu_3$ overtone band from \cite{burch_absorption_1969}, to guaranty the best adjustment with the latest version of HITRAN. We saw that even with the numerous updates that happened in HITRAN since 1989, only marginal corrections of the $\chi$ factors were sufficient to obtain an accurate match with experiment data. 
We also provide original $\chi$ factors for the $\nu_2$ band based on laboratory measurements from \cite{niro_spectra_2004} (see  left panel of \fig{fig:adj_CO2N2}). The factors were adjusted using data of the $\nu_2$ transition without small local bands (dotted lines in the left panel \fig{fig:adj_CO2N2}), then the temperature dependence was derived for different temperatures including local bands (solid colored lines in the left panel of \fig{fig:adj_CO2N2}). 
For the $\nu_1+\nu_3$ bands, \cite{burch_absorption_1969} give $\chi$ factors in a very different format than what we use, and without temperature dependence. For homogeneity and easy use, we re-determined $\chi$ factors following our formalism by using their experiment data (right panel of \fig{fig:adj_CO2N2}).
As measurements have been done at one single temperature, we assume that the temperature dependence is the same as the one of the $\nu_3$ band.
We notice that \cite{burch_absorption_1969} also provide measurements for the $\nu_3$ and $3\nu_3$ bands, that are in agreement with the more recent data we use. 

As shown in \fig{fig:adj_CO2N2}, our model based on the $\chi$ factor formalism is not able to accurately reproduce the profile of band centers at high pressures, both for the $\nu_2$ band and for weaker bands around 950 and 1050~cm$^{-1}$. This is due to subtle spectroscopic effects called line-mixing (see \sect{sec:discussion}) largely discussed in the literature \citep[e.g.][]{niro_spectra_2004, tran_measurements_2011, hartmann_chapter_2021-4}. Unfortunately, line-mixing modeling is extremely time-consuming, making it unadapted to compute the thousands of spectra required to generate a correlated-$k$ table. We consider that this localized difference, which would vanish at pressures of the order of a few bars or below, is acceptable in the context of climate modeling that is performing radiative transfer over the entire spectrum. A further analysis quantifying this error at various pressures could be useful to try to overcome this issue. 

\subsubsection{CO$_2$-H$_2$O} \label{sec:CO2H2O}
The only existing $\chi$ factors are for the $\nu_3$ band \citep{tran_measurements_2018}. To create a correlated-$k$ table usable for climate modeling, the spectra need to be computed down to a few tens of Kelvin. That is outside the validity domain given by \cite{tran_measurements_2018} (200-500~K). Moreover, we found that the B$_2$ coefficient, as formulated by \cite{tran_measurements_2018} (different from \eq{eq:B}) decreases below 150~K. This induces an inaccurate description of the far-wings below 100~K characterized by a non-monotonic temperature dependence. This induces an absorption increasing proportionally to the distance to the line-center at low temperature. For this reason, and to propose a common formulation of all the $\chi$ factors, we adapted them using the exponential formulation of \eq{eq:B}. To do so, we fit \eq{eq:B} on the analytical expression proposed by \cite{tran_measurements_2018} within the validity domain (200-500~K) to compute the temperature dependence ($\beta$ and $\gamma$ coefficients in Eq.~\ref{eq:B}). Then, we adjusted the $\chi$ factors on the laboratory measurements of the $\nu_3$ band from \cite{tran_measurements_2018} ($\alpha$ coefficient in Eq.~\ref{eq:B}). Thanks to this reformulation, the extrapolation of the $\chi$ factors seems coherent (i.e. decreasing absorption when increasing the distance from the line-center) down to few tens of Kelvin.

Other experiment data are provided by \cite{fleurbaey_characterization_2022} between 4200 and 4500~cm$^{-1}$ ($\nu_3$+$\nu_1$ band). 
Unfortunately, measurements down to 3700~cm$^{-1}$ are needed to accurately derive accurate $\chi$ factors for the $\nu_3$+$\nu_1$ band. 
We compared the spectrum of a H$_2$O+CO$_2$ gas mixture calculated with the $\chi$ factors we propose, with all measurement data from \cite{fleurbaey_characterization_2022} (see Fig.~8 therein) at 2860~cm$^{-1}$, near 4400~cm$^{-1}$, 5800~cm$^{-1}$ and 6500~cm$^{-1}$. Our results are in agreement with their conclusions, with a good match in most of the intervals except near 4400~cm$^{-1}$ for which we obtain the same difference between calculated spectrum and measurements.

Finally, regarding the intense absorption of the water continuum, the contribution of the $\nu_2$ band of CO$_2$ is negligible if CO$_2$ is not largely dominant (see \fig{fig:Earth} for instance). For these reasons, we consider that applying the CO$_2$-H$_2$O $\chi$ factors of the $\nu_3$ band over the entire spectrum is an acceptable assumption.

\subsection{The $\chi$ factors using HITEMP}\label{sec:HITEMP}

In this work, we derived the $\chi$ factors using HITRAN \cite{gordon_hitran2020_2021}, that is one of the most commonly used and up-to-date linelist in used in the exoplanet community to model temperate atmospheres. However, other databases such as HITEMP \citep{rothman_hitemp_2010} or ExoMol \citep{chubb_exomolop_2021} are more indicated to model the hot planets we currently more easily detect. The additional lines, which have small intensities at room temperature, present in these databases play an important role at high temperatures. We show in \fig{fig:bands_exp} the difference, in the infrared, between HITRAN (black line) and HITEMP (blue line) for a pure CO$_2$ gas at 1~bar and 750~K (that is the highest temperature for which we have experiment data). Horizontal gray lines indicate the intervals of available data experiments used to derive the $\chi$ factors. 
There is no significant difference between the databases at 750~K, in the wavenumber intervals where laboratory measurements are available.
For this reason, there is no need to derive specific $\chi$ factors from other linelists, and those proposed in this work can be used.

The future efforts for improving $\chi$ factor corrections for the benefit of the community should be focused on making additional measurements in spectral regions that are not constrained yet. Additionally, making more measurements at various temperatures could help constrain the temperature dependence, that is extremely important to model the wide variety of planets we detect. Characterizing ultra-short period rocky exoplanets such as 55-Cancri-e (e.g. CO$_2$ detection proposed by \citealt{hu_secondary_2024}) is challenging as they are largely warmer than laboratory measurement capabilities (few thousands Kelvin versus few hundreds). 
It is crucial to keep this point in mind when doing radiative transfer calculations of such environments. Regarding the difficulty of performing laboratory experiments at high temperature, this factual situation is likely not about to change. 

\subsection{New correlated-$k$ tables and continua}\label{sec:corrk}

Opacity data are often the limiting factor of climate modeling studies, in a sense that, due to the complexity of creating new correlated-$k$ tables for instance, it is challenging to model various atmospheric compositions. 
Based on the re-estimated $\chi$ factors we propose in this work, we computed several correlated-$k$ tables for different gas mixtures that are relevant for climate modeling of terrestrial planets, and freely accessible by the community at: \href{https://doi.org/10.5281/zenodo.15564548}{https://doi.org/10.5281/zenodo.15564548}. The high resolution spectra are calculated using \texttt{SpeCT} and the tables themselves are created with  \texttt{Exo\_k} \citep{leconte_spectral_2021}. Based on the conclusions of \cite{chaverot_how_2022} saying that inter-species molecular collisions induce a non-negligible contribution to the radiative balance of an atmosphere, we pay a particular attention to correctly model these processes, as described in \sect{sec:stateoftheart}.

We give new correlated-$k$ tables for 3 different gas mixtures: H$_2$O+CO$_2$, CO$_2$+N$_2$ and H$_2$O+CO$_2$+N$_2$, in a hdf5 format that is the regular format used by \texttt{Exo\_k}. All the tables are given at a resolution R=500\footnote{In \texttt{Exo\_k}, the resolution R is given in wavenumber as follows: $R=\Delta \sigma/B$ where $\Delta \sigma$ is the definition domain of the spectrum and $B$ the number of spectral intervals in the correlated-$k$ table.}, and include absorption lines from both CO$_2$ and H$_2$O. By using \texttt{Exo\_k}, it is easy for the user to decrease the resolution to make the tables usable in climate models. 
To deal with variable gas mixture compositions, all the tables include a volume mixing ratio (vmr) grid from 10$^{-6}$ to 1. For H$_2$O+CO$_2$ and CO$_2$+N$_2$, it corresponds to $P_{\rm H_2O}/P_{\rm total}$ and $P_{\rm CO_2}/P_{\rm total}$, respectively. For H$_2$O+CO$_2$+N$_2$ mixtures, one value of vmr is not sufficient to describe the gas composition. As these tables have been designed to model Earth like atmospheres, we fix the CO$_2$ vmr as a function of the dry atmosphere pressure (i.e. a table including 376~ppm of CO$_2$ means $P{_{\rm CO_2}}=376\times10^{-6}P{_{\rm N_2}}$). Therefore, the vmr grid in the table corresponds to $P_{\rm H_2O}/P_{\rm total}$. We provide 6 tables for this mixture of gases corresponding to various CO$_2$ concentrations: 376~ppm, 1000~ppm, 2000~ppm, 0.1, 0.5 and 0.75. For these 6 tables, the temperature range extends from 50~K to 1000~K, with pressures between 1~Pa and 10~bar. For H$_2$O+CO$_2$ and CO$_2$+N$_2$, the temperature range extends from 30~K to 2000~K, with pressures between 1~Pa and 100~bar.
An additional H$_2$O+N$_2$ correlated-$k$ table\footnote{\url{https://doi.org/10.5281/zenodo.5359157}} has been calculated in \cite{chaverot_how_2022} using an early version of \texttt{SpeCT}.

The correlated-$k$ tables contain only the line-centers contributions to the absorption coefficient (cut at $\pm$25~cm$^{-1}$) while the CIAs and continua are given separately.
We provide various original continua of CO$_2$, calculated using the updated $\chi$ factors described above, from 1~cm$^{-1}$ to 30000~cm$^{-1}$, and from 50~K to 3000~K at: \href{https://doi.org/10.5281/zenodo.15564548}{https://doi.org/10.5281/zenodo.15564548}. They correspond to CO$_2$ broadened by CO$_2$, N$_2$ or H$_2$O. The files are written in ascii and contain the wavenumbers (cm$^{-1}$) and the absorption in cm$^{-1}$.amagat$^{-2}$.

\section{Discussion}\label{sec:discussion}

\begin{figure*}[!ht]
    \centering\includegraphics[width=\linewidth]{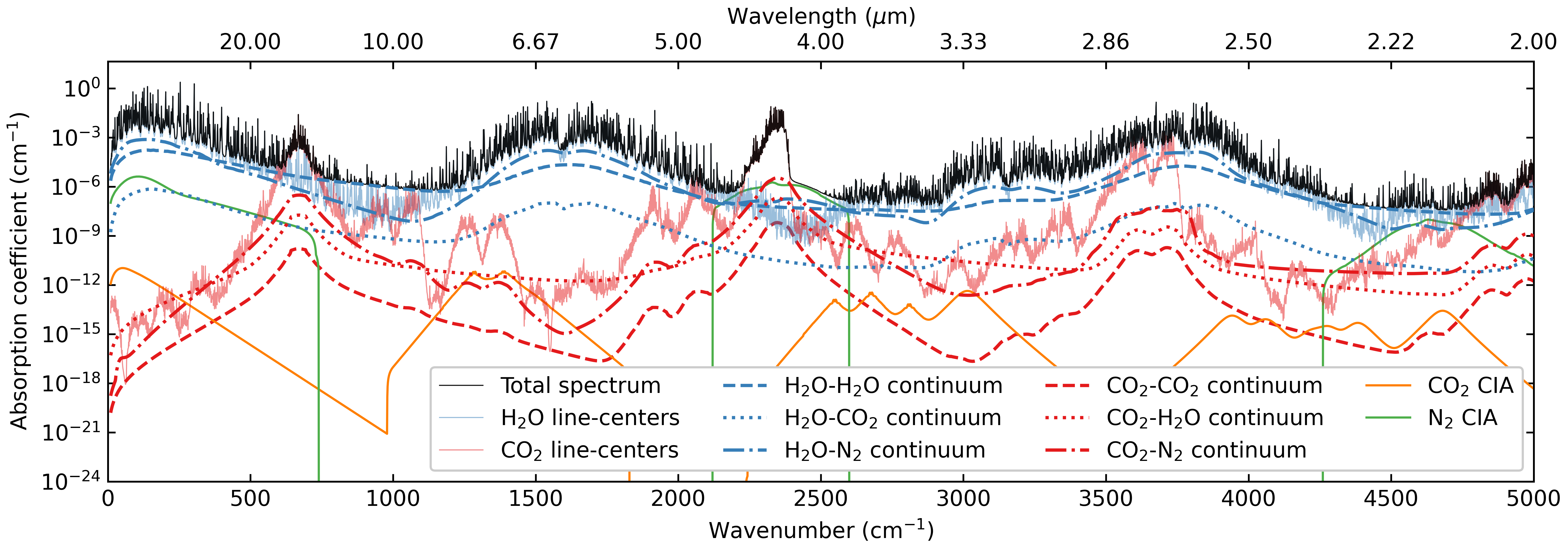}
    \caption{Absorption spectrum corresponding roughly to the surface Earth gas mixture (N$_2$, H$_2$O and CO$_2$). Here we assume 1~bar of total pressure, including 370~ppm of CO$_2$ and 0.01~bar of H$_2$O at 285~K. The CO$_2$ content corresponds to the Earth's atmosphere conditions in 2000.}
    \label{fig:Earth}
\end{figure*}

\begin{figure}[!ht]
    \centering\includegraphics[width=\linewidth]{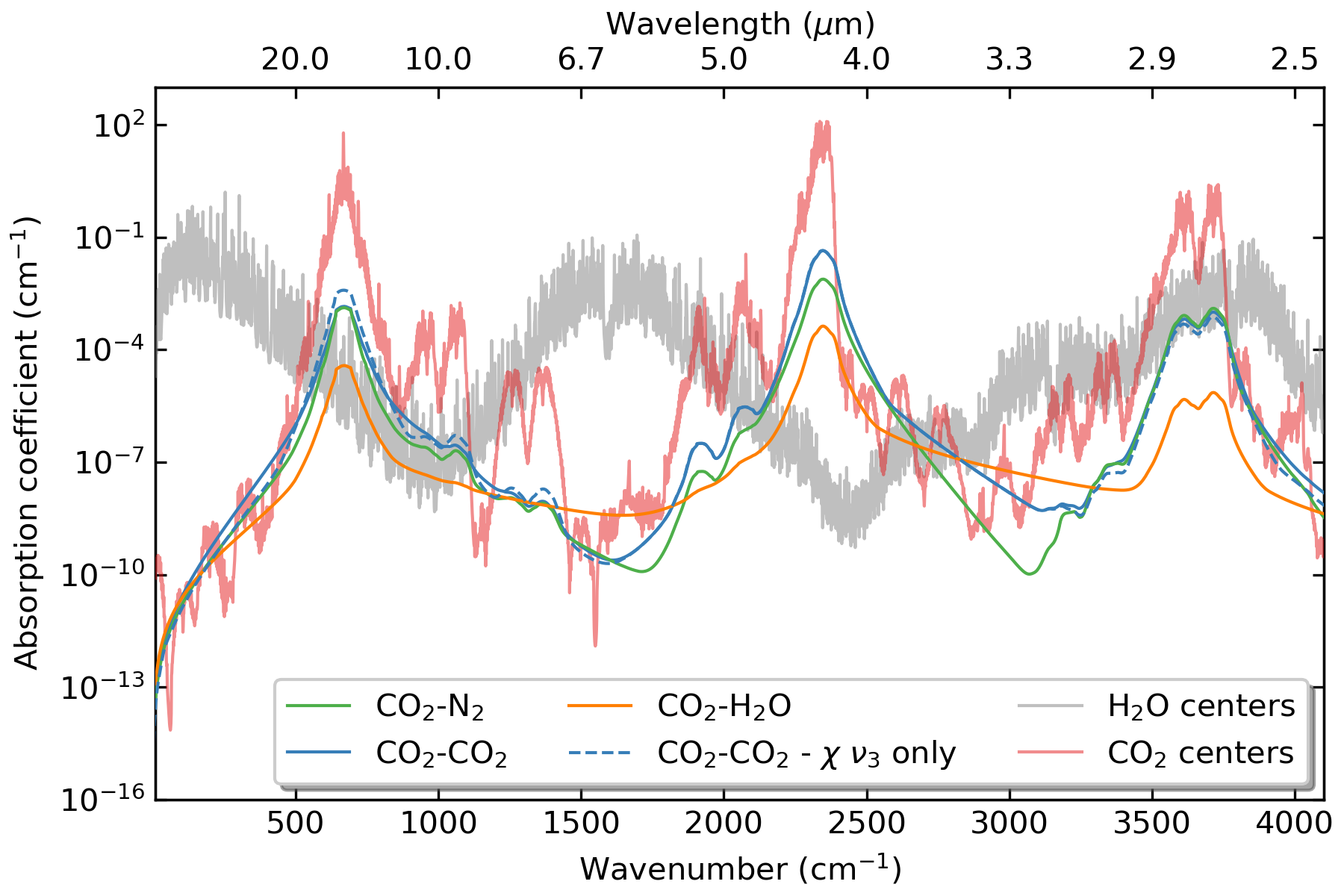}
    \caption{Absorption spectrum for a gas mixture including 0.01~bar of H$_2$O, 1~bar of CO$_2$ and 1~bar of N$_2$ at 300~K. The blue dashed line is the CO$_2$-CO$_2$ continuum obtained by applying the $\nu_3$ $\chi$ factor on the entire spectrum, while the solid blue line is the actual CO$_2$-CO$_2$ continuum. The red and gray lines are the CO$_2$ and H$_2$O line-centers, respectively. The green and orange lines are the CO$_2$-N$_2$ and CO$_2$-H$_2$O continua, respectively.}
    \label{fig:onlynu3}
\end{figure}

The different contributions to a total absorption spectrum are shown in \fig{fig:Earth}, for the atmospheric conditions at the Earth's surface. The total spectrum is the black line, while H$_2$O and CO$_2$ line-centers (from 0 to $\pm$25~cm$^{-1}$) are the blue and red lines, respectively. Continua (made of the far-wings) are in dashed, dotted and dash-dotted colored lines and CO$_2$ and N$_2$ CIAs are the orange and green lines respectively. 

As explained in \sect{sec:new_chi_factors}, for the CO$_2$-H$_2$O mixture, we use $\chi$ factors based on laboratory measurements of the $\nu_3$ band to correct the entire spectrum. As pointed out by \cite{fleurbaey_characterization_2022}, more measurements, especially for deriving the temperature dependence of the different contributions, are of first importance for climate modeling.
In order to have an idea of the error made by correcting the entire spectrum with one single set of $\chi$ factors, we calculated the absorption spectrum of CO$_2$-CO$_2$ (for which we have factors for the 3 main bands) by using the $\chi$ factors of the $\nu_3$ band only. A comparison with the accurate calculation is shown in \fig{fig:onlynu3} (blue lines). Even if the continuum calculated using only the $\chi$ factors of the $\nu_3$ band (dashed blue line) is qualitatively correct (with respect to a not corrected continuum, see \fig{fig:bands_exp}), some differences with the actual calculation remain. This is particularly true for the right side of the $\nu_3$+$\nu_1$ band, and for the core and wings of the small bands located on the right side of the $\nu_2$ band.

In order to follow what is usually done in the community, we use the formalism proposed by the MT\_CKD consortium, separating the far-wings from the line-centers at $\pm25$~cm$^{-1}$ from the core of the line. Even if they do not include $\chi$ factors in their calculations, \cite{gharib-nezhad_impact_2024} show that $\pm25$~cm$^{-1}$ is a good cutoff between the line-centers and the far-wings for pressures below 200~bar.
This empirical value is valid for Earth-like conditions, because it roughly separates the part of the line made of the line-center, from the part of the line made of far-wings (thus proportional to P$^2$). However, for very low and very high pressures, this assumption is not valid anymore. For instance, due to broadening, the core of the line (i.e. the part of the line proportional to P) may become much larger than 25~cm$^{-1}$ for high pressures. 
The numerical split between line-centers and continuum induces discontinuities in wavenumber in the continuum (made of the sum of the far-wings) at high pressures (1000~bars). This could induce errors due to the interpolation in climate models at such pressures. 

As discussed in \sect{sec:new_chi_factors}, a calculation based on the $\chi$ factor formalism is not able to accurately reproduce the shape of the core of the absorption bands at elevated pressure, for which the lines overlap significantly (see \fig{fig:adj_CO2N2}). This is a well known problem that can be solved by using a full line-mixing calculation \cite[e.g.][]{niro_spectra_2004, tran_measurements_2011}. 
At high pressure, collisions can induce transfers of populations, which lead to transfers of intensity between lines. These effects, called line-mixing, are also important at moderate pressures for the Q branch\footnote{Optical transitions for which the rotational quantum number in the ground state is the same as the rotational quantum number in the excited state} of the CO$_2$ $\nu_2$ band where lines significantly overlap.
The long computation time required by line-mixing calculations makes impossible the use of this method for creating correlated-$k$ tables including thousands of spectra. 
However, a mixed approach in which some high-pressure spectra would be calculated using a line-mixing calculation could be a great improvement on our work.

As shown by \fig{fig:bands_exp}, the amount of experimental data available to correct CO$_2$ spectra broaden by CO$_2$, H$_2$O or N$_2$ is limited. Moreover, there are not always measurements at multiple temperatures, making impossible any temperature constrain of the $\chi$ factors (e.g., $\nu_3$+$\nu_1$ band of CO$_2$-N$_2$ or $\nu_3$ band of CO$_2$-H$_2$O, see \tab{tab:LabRef}). The issue of CO$_2$-H$_2$O, for which there are measurements only for the $\nu_3$ band, cannot be resolved easily. Indeed, because of the strong absorption of the water continuum, even very small fractions of H$_2$O make the $\nu_2$ and $\nu_3$+$\nu_1$ bands undetectable, as shown by \fig{fig:Earth} reproducing roughly surface Earth conditions. 
Regarding temperature dependencies more specifically, no data exists beyond 770~K (see \tab{tab:LabRef}). To address community needs, we propose CO$_2$ continua up to 3000~K, calculated using temperature dependent chi factors fitted to experiments between 200~K and 770~K. Even if the extrapolation gives qualitatively correct continua, laboratory measurements are necessary to assess the accuracy of data extrapolated at a few thousand Kelvins. This point should be kept in mind when addressing the conclusions of studies on hot-Jupiters, for instance.

Regarding the major importance of opacity data for climate modeling, an experimental effort could be done making new experiments for pure CO$_2$, CO$_2$-N$_2$ and CO$_2$-H$_2$O in other spectral ranges, and/or at different temperatures. Making experiments and proposing plug-and-play corrections is a long scientific process, which needs to be prepared in advance to guaranty accurate and efficient tools able to interpret future observations of instruments in preparation. This modeling step, and the time required for it, should not be neglected.

\section{Conclusion}

In this work, we propose updated $\chi$ factors for CO$_2$ in different mixtures (pure CO$_2$, CO$_2$-N$_2$ and CO$_2$-H$_2$O), presented in \tab{tab:new_chi_factors}, which are needed to create accurate opacity data for climate modeling of terrestrial (exo)planets. These corrections coefficients are determined by adjusting synthetic spectra on laboratory measurements done at various pressure and temperature conditions. Consequently, the $\chi$ factor are empirical corrections that could need to be re-adjusted following the availability of new measurements or changes in the spectroscopic databases. Knowing this, we provide a complete and homogenized review of the relevant laboratory measurements available in the literature (\href{https://doi.org/10.5281/zenodo.15564294}{https://doi.org/10.5281/zenodo.15564294}) usable for future updates.

One of the main issue of modeling the climate of exoplanets is the creation of various opacity data covering the wide variety of potential atmospheric composition. To circumvent this issue, we give a set of 8 original correlated-$k$ tables (available at: \href{https://doi.org/10.5281/zenodo.15564548}{https://doi.org/10.5281/zenodo.15564548}, see \sect{sec:corrk} for a complete description of the data), based on the new $\chi$ factors, and calculated with \texttt{SpeCT} (described in \sect{sec:spect}) over a large range of temperatures, in the infra-red and optical.
We also propose a set of continua for pure CO$_2$, CO$_2$ broaden by N$_2$ or by H$_2$O, based on the formalism and methodology used by MT\_CKD for water. These continua, available at: \href{https://doi.org/10.5281/zenodo.15564548}{https://doi.org/10.5281/zenodo.15564548} are indicated to model the climate of planets dominated - or containing - CO$_2$. There are calculated from 1~cm$^{-1}$ to 30000~cm$^{-1}$, and from 50~K to 3000~K to cover all science cases.

\texttt{SpeCT}\footnote{\url{https://gitlab.com/ChaverotG/spect-public}} is a user-friendly open-source tool designed to calculate high resolution spectra, required to produce accurate correlated-$k$ tables. It is designed to easily add new gas mixtures and other $\chi$ factors corrections. We aim to continue developing \texttt{SpeCT} to cover a wider range of atmospheric compositions (i.e. by including CH$_4$ and O$_2$). The main difference between \texttt{SpeCT} and other codes from the literature \cite[e.g.][]{grimm_helios-k_2021} is that a particular effort is made for accurately modeling the pressure broadening from multi-species interactions. This makes \texttt{SpeCT} less versatile but more accurate. The effects of pressure broadening by different types of collision partner are particularly important when the considered gas mixtures are not dominated by one single gas \citep{chaverot_how_2022}, that is for intermediate mixing ratios. Consequently, the correlated-$k$ tables we propose are especially relevant for Earth-like atmosphere studies, or to model the runaway greenhouse.

As discussed in \sect{sec:discussion}, $\chi$ factors are an imperfect empirical correction of the modeling of spectra involving complex spectroscopic processes. For this reason, there are still some small differences between the actual measured spectrum and the numerical predictions. However, as the aim of this work if to produce opacity data, potential error are largely negligible in the final radiative transfer happening in climate models. 

\section*{Data Availability}
The correlated-$k$ tables and the CO$_2$ continua produced in this work are accessible at: \href{https://doi.org/10.5281/zenodo.15564548}{https://doi.org/10.5281/zenodo.15564548}. \texttt{SpeCT} is accessible at: \href{https://gitlab.com/ChaverotG/spect-public}{https://gitlab.com/ChaverotG/spect-public}. Finally, the laboratory measurements collected from the literature and used in this work are accessible at: \href{https://doi.org/10.5281/zenodo.15564294}{https://doi.org/10.5281/zenodo.15564294}.

\begin{acknowledgements}
GC acknowledges the financial support of the SNSF (grant number: P500PT\_217840). This work is supported by the French National Research Agency in the framework of the Investissements d'Avenir program (ANR-15-IDEX-02), through the funding of the "Origin of Life" project of the Grenoble-Alpes University. 
M.T. acknowledges support from the Tremplin 2022 program of the Faculty of Science and Engineering of Sorbonne University. M.T. acknowledges support from the High-Performance Computing (HPC) resources of Centre Informatique National de l'Enseignement Supérieur (CINES) under the allocations No. A0140110391 and A0160110391 made by Grand Équipement National de Calcul Intensif (GENCI). All authors thank the anonymous referee for their useful comments, which contribute to improving the manuscript.
\end{acknowledgements}

\bibliographystyle{aa}
\bibliography{biblio}

\appendix
\section{Overview of laboratory experiments data for CO$_2$ mixtures}\label{app:refs}

This section aim to address a non-exhaustive list of the most relevant laboratory measurements of the different CO$_2$ mixtures we consider. This bibliographic work has been used to derive the $\chi$ factors proposed in this article, and could be used to recalculate them following improvements of spectroscopic databases (see \tab{tab:LabRef}). 

\begin{table}[ht!]
	\centering
    \caption{List of the laboratory measurements available in the literature for the gas mixtures considered in this work.}
	\begin{tabular}{ c l l}
        \hline  \hline  
        Band & References & Temperatures (K)\\
        \hline 
        CO$_2$-CO$_2$& & \\
        $\nu_2$ & \cite{tran_measurements_2011} & 294; 373; 473 \\
        $\nu_3$ & \cite{tran_measurements_2011} & 218; 295; 473 \\
                & \cite{hartmann_measurements_1989} & 291; 414; 534\\
                & & 627; 751 \\
                & \cite{doucen_temperature_1985} & 193; 218; 238;\\ & & 258; 296 \\
                & \cite{burch_absorption_1969} & 295 \\
        $\nu_3$+$\nu_1$ & \cite{mondelain_co2_2017-1} & \\
                & \cite{tran_measurements_2011} & 230; 260; 295; 373 \\
                & \cite{tonkov_measurements_1996} & 295 \\
                & \cite{burch_absorption_1969} & 295 \\
        $\approx$5900~cm$^{-1}$& \cite{kassi_high_2015} & 293 \\
        3$\nu_3$ & \cite{burch_absorption_1969} & 295 \\
        \hline
        CO$_2$-N$_2$ & & \\
        $\nu_2$ & \cite{niro_spectra_2004} & 198; 255; 273; 295 \\
                & \cite{cousin_temperature_1985} & 193; 218; 238; 296 \\
        $\nu_3$ & \cite{perrin_temperature-dependent_1989} & 296; 448; 550; \\
                &                                          & 623; 643; 773\\
                & \cite{burch_absorption_1969} & 295 \\
        $\nu_3$+$\nu_1$ & \cite{burch_absorption_1969} & 273 \\
        3$\nu_3$ & \cite{burch_absorption_1969} & 295 \\
        \hline
        CO$_2$-H$_2$O & & \\
        $\nu_3$ & \cite{tran_measurements_2018} & 325-367  \\
                & \cite{baranov_significant_2016} &  295-339 \\
        $\nu_3$+$\nu_1$ & \cite{fleurbaey_characterization_2022} & 293\\ 
        $\approx$5800~cm$^{-1}$ & \cite{fleurbaey_characterization_2022} & 293 \\
        $\approx$6500~cm$^{-1}$ & \cite{fleurbaey_characterization_2022} & 293 \\
	\end{tabular}
 \label{tab:LabRef}
\end{table}

For the CO$_2$-CO$_2$ gas mixture, the most recent and complete data are from \cite{tran_measurements_2011} for the three main bands of CO$_2$: $\nu_2$, $\nu_3$ and $\nu_1+\nu_3$. Measurements of the $\nu_3$ band are also available in \cite{burch_absorption_1969}, \cite{doucen_temperature_1985} and \cite{hartmann_measurements_1989}. They all propose $\chi$ factors, based on different formalisms. The $\nu_3$ band should be corrected together with measurements of the $3\nu_3$ band ([7000;7100]~cm$^{-1}$) from \cite{burch_absorption_1969} and \cite{filippov_collision-induced_1997}. Finally, other measurements of the $\nu_1+\nu_3$ band are proposed in \cite{burch_absorption_1969} and \cite{tonkov_measurements_1996}.
Measurements have been done for pure CO$_2$ by \cite{kassi_high_2015} and \cite{mondelain_co2_2017-1} near 5900~cm$^{-1}$ (see \fig{fig:bands_exp}) but they are not used in this work.

For the CO$_2$-N$_2$ gas mixture, the only available measurements for the $\nu_2$ band are from \cite{niro_spectra_2004}. Measurements of the $\nu_3$ band as been done by \cite{perrin_temperature-dependent_1989} and \cite{cousin_temperature_1985}. They both proposed $\chi$ factors, but we based our work on the most recent measurements from \cite{perrin_temperature-dependent_1989}. Additionally, \cite{burch_absorption_1969} provide measurements of the $\nu_3$ and $3\nu_3$ ([7000;7100]~cm$^{-1}$) bands, that should be used together to adjust $\chi$ factors.
They also give measurements of the $\nu_1+\nu_3$ band.

For the CO$_2$-H$_2$O gas mixture, for the reasons explained in \sect{sec:new_chi_factors}, there are not many measurements of the $\nu_2$ and $\nu_1+\nu_3$ bands. For the $\nu_3$ bands, the most recent data are from \cite{tran_measurements_2018}. Previous measurements done by \cite{baranov_significant_2016} are discussed in \cite{tran_measurements_2018}.
Other measurements have been done for CO$_2$-H$_2$O by \cite{fleurbaey_characterization_2022} near 6500~cm$^{-1}$ (see \fig{fig:bands_exp}) but they are not used in this work.

\end{document}